\documentclass[lettersize,journal]{IEEEtran}

\usepackage{amsmath,amsfonts}
\usepackage{amssymb}
\usepackage{array}
\usepackage[caption=false,font=normalsize,labelfont=sf,textfont=sf]{subfig}
\usepackage{textcomp}
\usepackage{stfloats}
\usepackage{url}
\usepackage{verbatim}
\usepackage{graphicx}
\usepackage{cite}
\usepackage{xcolor}
\usepackage{booktabs}
\usepackage{multirow}
\usepackage{makecell}
\usepackage{tabularx}
\usepackage{enumerate}
\usepackage{enumitem}
\usepackage{microtype}
\usepackage{balance}
\usepackage{bm}
\usepackage{ntheorem}
\usepackage[linesnumbered,algoruled,boxed,lined,noend]{algorithm2e}

\theorembodyfont{}

\allowdisplaybreaks[4]

\begin{document}

\title{Recovery of UAV Swarm-enabled Collaborative Beamforming in Low-altitude Wireless Networks under Wind Field Disturbances}

\author{Geng Sun,~\IEEEmembership{Senior Member,~IEEE,}
        Chenbang Liu,
        Jiahui Li,
        Guannan Qu,
        Shuang Liang,
        Jiacheng Wang,
        Changyuan Zhao,
        Dusit Niyato,~\IEEEmembership{Fellow,~IEEE}

        \thanks{
        \par This study is supported in part by the National Natural Science Foundation of China (62272194, 62471200), in part by the Science and Technology Development Plan Project of Jilin Province (20250101027JJ), in part by the Postdoctoral Fellowship Program of China Postdoctoral Science Foundation (GZC20240592), in part by China Postdoctoral Science Foundation General Fund (2024M761123), and in part by the Scientific Research Project of Jilin Provincial Department of Education (JJKH20250117KJ). (\emph{Corresponding author: Jiahui Li.})
        
        \par Geng Sun is with the College of Computer Science and Technology, Key Laboratory of Symbolic Computation and Knowledge Engineering of Ministry of Education, Jilin University, Changchun 130012, China, and also with the College of Computing and Data Science, Nanyang Technological University, Singapore 639798 (e-mail: sungeng@jlu.edu.cn).
        
        \par Chenbang Liu, Jiahui Li, and Quannan Qu are with the College of Computer Science and Technology, Jilin University, Changchun 130012, China (e-mails: liucb23@jlu.edu.cn; lijiahui@jlu.edu.cn;  gnqu@jlu.edu.cn).
        
        \par Shuang Liang is with the School of Information Science and Technology, Northeast Normal University, Changchun, 130117, China, and also with Key Laboratory of Symbolic Computation and Knowledge Engineering of Ministry of Education, Jilin University, Changchun 130012, China (e-mail: liangshuang@nenu.edu.cn).

        \par Jiacheng Wang and Dusit Niyato are with the College of Computing and Data Science, Nanyang Technological University, Singapore (e-mails: jiacheng.wang@ntu.edu.sg; dniyato@ntu.edu.sg).
        
        \par Changyuan Zhao is with the College of Computing and Data Science, Nanyang Technological University, Singapore, and CNRS@CREATE, 1 Create Way, 08-01 Create Tower, Singapore 138602 (e-mail: zhao0441@e.ntu.edu.sg).

        }

        
        

        }

\markboth{IEEE Internet of Things Journal,~Vol.~XX, No.~X, XXXX~2025}%
{Sun \MakeLowercase{\textit{et al.}}: Recovery of UAV Swarm-enabled Collaborative Beamforming under Wind Field Disturbances}

\IEEEpubid{0000--0000/00\$00.00~\copyright~2025 IEEE}

\maketitle

\begin{abstract}
Unmanned aerial vehicle (UAV) swarms utilizing collaborative beamforming (CB) in low-altitude wireless networks (LAWN) demonstrate significant potential for enhanced communication range, energy efficiency, and signal directivity through the formation of virtual antenna arrays (VAA). However, environmental disturbances, particularly wind fields, significantly degrade CB performance by introducing positional errors that disrupt beam patterns, thereby compromising transmission reliability. This paper investigates the critical challenge of maintaining CB performance in UAV-based VAAs operating in LAWN under wind field disturbances. We propose a comprehensive framework that models the impact of three distinct wind conditions (constant, shear, and turbulent) on UAV array performance, and formulate a long-term real-time optimization problem to maximize directivity while minimizing maximum sidelobe levels through adaptive excitation current weight adjustments. To address the inherent complexity of this problem, we propose a novel proximal policy optimization algorithm with long short-term memory (LSTM) structure and adaptive learning rate (PPO-LA), which effectively captures temporal patterns in wind field disturbances and enables real-time adaptation without requiring extensive prior training for specific wind conditions. Our simulation results demonstrate that the proposed PPO-LA algorithm successfully recovers degraded CB performance across various wind scenarios, and thus significantly outperforming benchmark algorithms.
\end{abstract}


\section{Introduction} 
\label{sec:introduction}

\IEEEPARstart{U}{nmanned} aerial vehicles (UAVs) have gained significant momentum in both civil and military applications due to their high mobility, low cost, and versatility~\cite{Wang2022,Song2022Multi-UAV,Masum2013Simulation,Wang2025Toward}. Recently, the deployment of multiple UAVs as a swarm enables cooperative execution of complex missions in low-altitude wireless networks (LAWN) \cite{yuan2025groundskyarchitecturesapplications,li2025securingskyintegratedsatelliteuav} that would be impossible for individual vehicles~\cite{Wu2018Joint}. UAV swarms effectively overcome the inherent limitations of a single UAV in terms of energy capacity and coverage range. In particular, by introducing collaborative beamforming (CB) method, UAV swarms can form virtual antenna arrays (VAA) to achieve remarkable potential for obtaining maximum gain in the intended direction \cite{Zheng2025UAV}. Consequently, CB enables independent nodes to collaborate, thereby enhancing transmission distance, energy efficiency, and signal directivity~\cite{Jayaprakasam2017}.

\IEEEpubidadjcol

\par Despite the theoretical advantages of CB in UAV swarms \cite{He2025Satellite}, practical implementation faces critical challenges from environmental disturbances, particularly those caused by wind conditions. Even minor positional displacements of UAVs can significantly degrade beam patterns and transmission performance~\cite{Feng2021Positional}. This degradation is further exacerbated by error propagation in relative formations within the swarm~\cite{Du2024Location}, which leads to substantial deterioration in communication reliability and efficiency. Traditional approaches to address these challenges often rely on pre-programmed responses or static algorithms that fail to adapt to dynamic environmental conditions in real-time~\cite{Padakandla2020}.

\par Recent methods have attempted to exploit swarm intelligence in UAV swarms to address disturbance-induced performance degradation~\cite{Wan2023An}. However, these implementations still face significant limitations that restrict their practical application. Existing approaches struggle with real-time adaptation due to two fundamental constraints. First, such methods depend on extensive training data for specific wind profiles, which makes them ineffective in new or changing conditions. Second, these methods are difficult to handle continuous variations in disturbance patterns. These limitations are particularly problematic in dynamic real-world environments where wind conditions can change rapidly and unpredictably.

\par To address these limitations, this paper investigates a novel approach capable of compensating for wind-induced performance losses in CB-based UAV swarms operating in LAWN in real-time. Previous research has demonstrated the effectiveness of online joint optimization approaches in UAV swarm systems for various applications \cite{He2024QoE}, which motivates our development of adaptive algorithms for dynamic wind conditions. However, developing such a solution presents several significant challenges. First, modeling the complex interactions between wind field disturbances and UAV positional stability requires capturing non-linear aerodynamic effects that vary across different UAV configurations and environmental conditions. Second, formulating an appropriate optimization problem is complicated by the inherent trade-offs between maintaining beam directivity and minimizing side lobe levels while subject to constantly changing disturbance patterns. Third, ensuring real-time adaptability necessitates algorithms that can process environmental feedback and implement compensatory adjustments within strict latency constraints. 

\par Accordingly, we seek to model the wind field disturbance and propose an online algorithm to overcome the issue, and thus offer a framework for continuous environmental interaction and policy improvement without requiring extensive prior training data for specific conditions. The main contributions of this paper are summarized as follows:

\begin{itemize}
    \item \textit{Wind Field Disturbance Modeling and Problem Formulation}: We establish a framework for UAV array performance degradation under three distinct wind conditions (constant, shear, and turbulent). Following this, we formulate a long-term and real-time optimization problem for maximizing the directivity and minimizing the maximum sidelobe level of the UAV-based VAAs by adjusting the excitation current weights of the UAVs.

    \item \textit{Real-Time Deep Reinforcement Learning (DRL) Recovery}: We propose a novel proximal policy optimization (PPO) with long short-term memory (LSTM) structure and an adaptive learning rate (PPO-LA) algorithm. In this algorithm, the incorporated LSTM structure and adaptive learning rate method can enhance policy networks for superior temporal pattern recognition.

    \item \textit{Performance Evaluation}: Simulation results show that the proposed methods can recover the degraded CB performance. Moreover, the PPO-LA can adapt to dynamic disturbances and outperforms various benchmark algorithms. 
\end{itemize}

\par The remainder of this paper is organized as follows. Section \ref{sec:related_work} reviews the related works. Section \ref{sec:models_and_preliminaries} presents the models and preliminaries. Section \ref{sec:algorithm} proposes the algorithm. Section \ref{sec:simulation_results_and_analysis} shows the simulation results and Section \ref{sec:conclusion} concludes the paper.

%
%
\section{Related Work} %
\label{sec:related_work}

\par In this section, we review the existing literature on CB in UAV networks and error recovery techniques for UAVs under disturbances to highlight our innovations and contributions. The comparisons between these related works and our work are shown in Table \ref{comparison} and Table \ref{comparison1}.

%
%
\subsection{Performance Analysis of CB in UAV Networks}

\par CB for UAV networks has been extensively studied to enhance communication performance through various approaches and optimization techniques. Izydorczyk \textit{et al.} \cite{Izydorczyk2018} demonstrated the superiority of CB over maximum ratio combining receivers in UAV communications over long-term evolution networks. Jung \textit{et al.} \cite{Jung2022} proposed a dynamic random VAA architecture to enhance secrecy energy efficiency under eavesdropping attacks, and this architecture highlights the importance of phase error compensation. Sun \textit{et al.} \cite{Sun2021} developed a limited-feedback CB method, which achieves higher signal-to-noise ratios and extended coverage ranges in mobile UAV networks. Amer \textit{et al.} \cite{Amer2019} introduced a dual-domain CB scheme, which serves both aerial and ground users in dense networks. Mei \textit{et al.} \cite{Mei2020} designed an interference suppression method for UAV-ground coexistence scenarios through coordinated beam nulling techniques.

\par Recent studies have advanced CB techniques for UAV networks through innovative approaches. Han \textit{et al.} \cite{Han2025} proposed a 3D sidelobe suppression algorithm that enhances interference management by dynamically selecting subsets of UAVs to form beams. Ouassal \textit{et al.} \cite{Ouassal2021} introduced a decentralized frequency alignment method, which demonstrates robust synchronization in distributed phased arrays under dynamic network conditions. Jayaprakasam \textit{et al.} \cite{Jayaprakasam2017} surveyed distributed beamforming in wireless sensor networks, which emphasizes energy-efficient strategies for aerial-ground co-existence. Wan \textit{et al.} \cite{Wan2024} integrated CB with cell-free radio access networks using network-assisted full-duplex, and this integration shows improved spectral efficiency for cellular-connected UAVs. These works collectively address key challenges in scalability, interference control, and system integration for UAV-assisted communications.

\par Despite these advancements, current CB research predominantly assumes ideal operational conditions, and thus often neglecting performance degradation caused by UAV position disturbances. Existing studies typically rely on perfect array geometry or static deployment scenarios, while they neglect dynamic disturbances. As demonstrated by Gong \textit{et al.} \cite{Gong2017}, channel uncertainties in relay networks can significantly degrade beamforming performance when structural information about the distribution is ignored. Jung \textit{et al.} \cite{Jung2019} further reveal that even minor phase offsets in virtual antenna arrays, caused by position estimation errors, critically impair secrecy rates in collaborative beamforming systems. These limitations highlight a fundamental gap in CB research, as real-world deployments must account for dynamic perturbations that disrupt phase synchronization, a prerequisite for optimal beamforming performance.

%
%
\subsection{Error Recovery in UAVs Under Disturbances}

\par The impact of wind fields on UAV operational stability has been widely recognized in recent studies.  Zhang and Yan \cite{Zhang2021} demonstrated that wind disturbances can cause significant steady-state errors in fixed-wing UAV formations, which requires advanced control strategies to maintain geometric configuration. Wang \textit{et al.} \cite{Wang2021} revealed through detailed aerodynamic modeling that wind gusts create complex excitation patterns across joined-wing UAV structures, which leads to substantial attitude deviations that challenge conventional control systems. For quadrotor platforms, Jeon \textit{et al.} \cite{Jeon2021} established that wind effects can compromise system stability beyond certain thresholds, and the critical conditions depend on specific vehicle dynamics.

\par Recent developments in disturbance mitigation have incorporated advanced modeling techniques. Sun \textit{et al.} \cite{Sun2020} employed artificial neural networks trained with experimental wind data to predict and counteract wake effects, which achieves improved power recovery through optimized control strategies. Bisheban \textit{et al.} \cite{Bisheban2021} introduced geometric adaptive controllers with neural network augmentation that maintain stable tracking performance under wind disturbances without requiring prior knowledge of wind conditions. De Simone \textit{et al.} \cite{Marco2017} implemented refined aerodynamic models in simulation environments to better characterize thrust variations during wind disturbances, and thus showing that simplified assumptions may underestimate control requirements.

\par Various error recovery techniques have been proposed to enhance UAV operational reliability under environmental disturbances and system errors \cite{Chen2025Joint}. Shi \textit{et al.} \cite{Shi2018} proposed a software-defined radio platform to recover real-time CSI errors in UAV beamforming systems, which validates the necessity of robust error correction for mobile aerial networks. Patnaik \textit{et al.} \cite{Patnaik2021} developed a foldable quadrotor system with adaptive morphology control to mitigate positioning errors and environmental disturbances during autonomous missions. He \textit{et al.} \cite{He2019} designed an online subspace tracking algorithm to detect and correct sensor anomalies in UAV flight data, which provides a computational framework for real-time error recovery. Mohammadi \textit{et al.} \cite{Alireza2020} proposed a vision-aided model predictive control scheme to compensate for positional deviations during UAV landing procedures, which demonstrates the effectiveness of multisensory fusion in wind disturbance rejection.

\par Despite these advances, current error recovery methods have significant limitations when applied to CB with UAV swarms. Most existing approaches focus solely on single-UAV position correction without considering the collective array geometry required for CB, or they require extensive pre-training for specific disturbance patterns, which limits adaptability to dynamic wind conditions. Additionally, these methods typically involve high computational complexity that challenges the real-time performance requirements of UAV swarms communications \cite{Zhang2025Covert}, and they rarely address the unique phase synchronization challenges critical to maintaining beamforming integrity under position disturbances. 

\par Unlike previous works that address individual UAV error recovery or idealized CB models, our research investigates the specific challenges of maintaining CB performance in UAV-based VAAs under relatively realistic wind field disturbances, which provides compensation mechanisms for wind-induced array position errors.

%
%
\section{Models and Preliminaries} 
\label{sec:models_and_preliminaries}

\par In this section, we first present the system overview. Subsequently, we introduce the adopted models in detail, including the VAA model, communication model, and physical environment wind field model. The main symbols used in this paper are shown in Table \ref{tab:notations}.

\subsection{System Overview}
\label{subsec:system_overview}

\par
\begin{figure}[tb]
  \centering
  \includegraphics[width=3.5in]{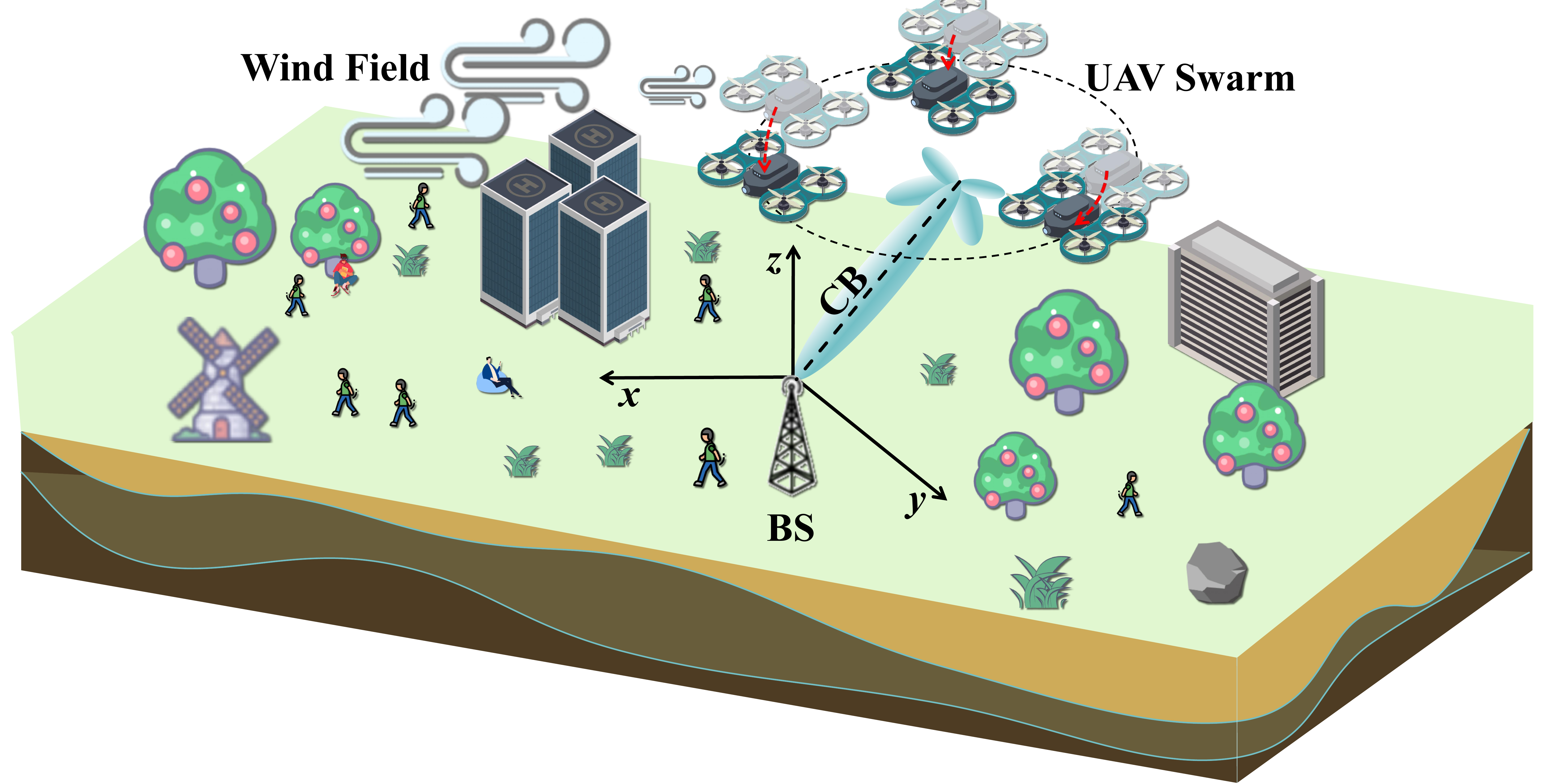}
  \caption{Communication between the UAV swarm and a BS of LAWN in a wind field.}
  \label{fig:network-model}
\end{figure}

 As shown in Fig.~\ref{fig:network-model}, we consider a CB-based UAV communication system in LAWN operating within a physical wind field. In this system, a UAV swarm consisting of $N$ UAVs, denoted as $\mathcal{U}$, are randomly distributed in three-dimensional (3D) space, and maintain communication with a ground base station (BS) denoted as $\Gamma$. In this case, the UAV swarm communicates with $\Gamma$  using CB. However, in such a high-altitude environment, the presence of wind fields may influence the physical 3D positions of the UAVs.

\par In such a dynamic environment, the transmission performance of the UAV-based VAA may be degraded because of the effect of the wind field on UAV positions. To mitigate these disturbances, the UAV-based VAA can dynamically adjust the transmission parameters of the UAVs, such as excitation current weights, according to the current positions of the UAVs at each timeslot. 

\par Without loss of generality, we consider a 3D Cartesian coordinate system with a discrete-time system evolving over timeline $\mathcal{T} = \{t|t = 1,2,\ldots, T \}$ to model the considered scenario mathematically. Consequently, the position and excitation current weight of the $i$-th UAV at timeslot $t$ are given by $(x_{i,t}^{\mathcal{U}}, y_{i,t}^{\mathcal{U}}, z_{i,t}^{\mathcal{U}})$ and $\omega_{i,t}^{\mathcal{U}}$, respectively. Additionally, the BS $\Gamma$ is located at a fixed point $(x^{\Gamma}, y^{\Gamma}, z^{\Gamma})$. 

\par In the following, we will introduce three key models, which are the VAA model, communication model, and wind field model.

\subsection{Virtual Antenna Array Model}
\label{subsec:virtual_antenna_array_model}

\par VAA model leverages the collective capabilities of multiple UAV antennas by using CB to enhance signal directionality. Specifically, the electromagnetic waves radiated by UAV antennas within the VAA interact to form a composite beam pattern, which is characterized by a prominent main lobe and reduced sidelobes. To quantify the strengths of the signals in different directions, we introduce the array factor index, which is given by

\begin{equation}
\label{eq:AF}
\begin{aligned}
AF_t(\theta_d, \phi_d) = \sum_{i=1}^{N} \omega_{i,t}^{\mathcal{U}} 
&\exp\big[jk(x_{i,t}^{\mathcal{U}} \sin\theta_d \cos\phi_d \\
&+ y_{i,t}^{\mathcal{U}} \sin\theta_d \sin\phi_d \\
&+ z_{i,t}^{\mathcal{U}} \cos\theta_d)\big],
\end{aligned}
\end{equation}

\noindent where $j$ is the imaginary unit, $k=2\pi/\lambda$ is the free-space wave number, $\lambda$ is the signal wavelength, and $\theta_d \in [0, \pi]$, $\phi_d \in [-\pi, \pi]$ represent the elevation and azimuth angles, respectively. Moreover, the $\omega_{i,t}^{\mathcal{U}}$ represents the excitation current weight of the $i$-th UAV at timeslot $t$.

\par Clearly, the positions and excitation current weights of the UAVs play a vital role in determining the optimal radiation pattern of the VAA model. Therefore, we can fine-tune the excitation current weights of the UAVs to ensure communication efficiency, when the positions of the UAVs are blown off course by the wind.

{\footnotesize
\begin{table*}[t]
\centering
\caption{\textsc{Comparison of CB Techniques in UAV Networks}}
\label{comparison}
\setlength{\tabcolsep}{2pt}
\renewcommand{\arraystretch}{1}
\begin{tabular}{|p{0.1\textwidth}|p{0.1\textwidth}|p{0.18\textwidth}|p{0.25\textwidth}|p{0.15\textwidth}|p{0.2\textwidth}|}
\hline
\textbf{Related works} & \textbf{Number of UAVs} & \textbf{Communication scenarios} & \textbf{Optimization variables} & \textbf{Network optimization metrics} & \textbf{Optimization methods} \\
\hline
\cite{Izydorczyk2018} & Single UAV & Long term evolution (LTE) networks with interference & Beam selection strategies (power, reference signal received power, reference signal received quality, signal-to-interference-plus-noise ratio (SINR)) & SINR performance, outage probability & Comparative analysis between CB and maximum ratio combining (MRC) techniques \\
\hline
\cite{Jung2022} & UAV swarm & Secure air-to-ground communication with eavesdroppers & VAA topology randomization, phase compensation & Secrecy rate, security energy efficiency & Analog CB with stochastic VAA \\
\hline
\cite{Sun2021} & 8-16 UAVs & Remote base station (BS) communication with energy constraints & 3D positions, flight speeds, excitation weights & Transmission time, performing time, energy consumption & Improved multi-objective ant lion optimization (IMOALO) \\
\hline
\cite{Amer2019} & Single UAV & Ground and aerial user co-existence in cellular networks & Antenna down-tilt angle, number of scheduled users, number of antennas & Successful content delivery probability, spectral efficiency & Conjugate beamforming with interference coordination \\
\hline
\cite{Mei2020} & Multiple UAVs (cellular-connected) & Downlink communication in cellular network with co-existing UAVs and ground user equipment & Power allocations at serving BS for UAV and messages of terrestrial user equipment & SINR performance & Centralized optimization and distributed divide-and-conquer approach \\
\hline
\cite{Han2025} & Multiple UAVs & 3D networks with desired and undesired receivers & UAV selection for CB, phase alignment & Interference control at undesired receivers, array gain & Sidelobe suppression algorithm with UAV selection \\
\hline
\cite{Ouassal2021} & Multiple nodes & Distributed phased arrays with frequency drift & Frequency alignment, phase coordination & Coherent gain, synchronization accuracy & Decentralized frequency consensus protocol \\
\hline
\cite{Jayaprakasam2017} & Wireless sensor nodes & General wireless sensor network applications & Node selection, power allocation & Energy efficiency, network lifetime & Classification of distributed CB approaches \\
\hline
\cite{Wan2024} & Multiple UAVs & Cell-free radio access network with full-duplex & Beamforming weights, access point clustering & Uplink/downlink rates, spectral efficiency & Deep Q-network optimization \\
\hline
\cite{Gong2017} & Device-to-device relay nodes & Underlay cellular networks & Beamforming weights, power allocation & Secrecy rate, interference control & Distributionally robust optimization \\
\hline
\cite{Jung2019} & VAA elements & Secure communication with eavesdroppers & Phase adjustment, VAA topology & Secrecy rate, array gain & Open-loop analog CB with phase error analysis \\
\hline
\end{tabular}
\end{table*}
}

{\footnotesize
\begin{table*}[t]
\centering
\caption{\textsc{Comparison of Error Recovery Techniques for UAVs Under Disturbances}}
\label{comparison1}
\setlength{\tabcolsep}{3pt}
\renewcommand{\arraystretch}{1}
\begin{tabular}{|p{0.1\textwidth}|p{0.1\textwidth}|p{0.18\textwidth}|p{0.25\textwidth}|p{0.15\textwidth}|p{0.2\textwidth}|}
\hline
\textbf{Related works} & \textbf{Number of UAVs} & \textbf{Scenarios} & \textbf{Error Type Addressed} & \textbf{Technical Approach} & \textbf{Disturbance Type} \\
\hline
\cite{Zhang2021} & 3 UAVs & Formation flight of fixed-wing UAVs in wind conditions & Position tracking errors caused by wind-induced disturbances & Robust proportional-integral-derivative controller with integrated disturbance rejection & Continuous wind fields with varying intensity \\
\hline
\cite{Wang2021} & Single UAV & High-altitude flight of joined-wing UAV configuration & Pitch angle oscillations due to traveling wave excitations & High-fidelity distributed aerodynamic modeling approach & 3D turbulent wind gusts \\
\hline
\cite{Jeon2021} & Single UAV & Quadrotor stabilization in windy environments & Angular velocity instability from aerodynamic moments & Linear quadratic regulator control with wind effect compensation & Directional wind disturbances affecting rotor dynamics \\
\hline
\cite{Sun2020} & 5 UAVs & Coordinated operation of UAVs in wind farm & Power output reduction from wake interference & Artificial neural network (ANN)-based wake prediction and compensation & Downstream wake effects between multiple UAVs \\
\hline
\cite{Bisheban2021} & Single UAV & Aggressive quadrotor maneuvers under wind conditions & Trajectory tracking errors and attitude deviations & Geometric adaptive control with neural networks & Sudden wind gusts from varying directions \\
\hline
\cite{Marco2017} & Single UAV & Quadrotor flight simulation with wind effects & Thrust generation errors during wind disturbances & Blade element theory for thrust modeling & Crosswind and headwind conditions \\
\hline
\cite{Shi2018} & Single UAV & Air-to-ground communication under mobility conditions & Channel state information delays and reciprocity errors & Beamforming with explicit channel state information feedback and extended Kalman filter (EKF) & Wireless multipath fading and Doppler effects during movement \\
\hline
\cite{Patnaik2021} & Single UAV & High-impact collisions during cluttered environment navigation & Post-impact rebound instability and structural damage & Foldable quadrotor with torsional springs and recovery controller & Physical collisions with static/dynamic obstacles at various velocities \\
\hline
\cite{He2019} & Single UAV & Real-time flight data monitoring during operation & Multivariate sensor anomalies and outlier measurements & Online subspace tracking for anomaly detection and mitigation & High-frequency sensor noise and heterogeneous data corruption\\
\hline
\cite{Alireza2020} & Single & Precision landing on moving platforms outdoors & Vision-based localization drift and tracking inaccuracies & Model predictive control with gimbaled camera and EKF fusion & Wind gusts and unpredictable platform motion dynamics \\
\hline
\end{tabular}
\end{table*}
}

\subsection{Communication Model}
\label{subsec:communication_model}

\par In this part, we present the communication model for the UAV-enabled system. We consider that the UAVs within the VAA have an isotropic element pattern, which ensures uniform radiation characteristics. Following this, let $(\theta_d^{\mathcal{U}}, \phi_d^{\mathcal{U}})$ represent the direction from the center of the UAV to the receiver. Subsequently, the directivity $D$, which is the key performance metric that quantifies the concentration of radiated power in the intended direction, is given by
\begin{equation}
  \label{eq:DIR}
  D = \frac{4\pi P_t (\theta_d^{\mathcal{U}}, \phi_d^{\mathcal{U}})}{\int_{\theta_d=0}^{\pi} \int_{\phi_d=-\pi}^{\pi}P_t(\theta_d, \phi_d)\sin\theta_d d\theta_d d\phi_d},
\end{equation}

\noindent where $P_t(\theta_d, \phi_d)$ denotes the radiated power in the direction $(\theta_d, \phi_d)$. Moreover, the maximum sidelobe level $M$, a critical metric for suppressing interference from unintended directions, is expressed as follows:
\begin{equation}
  \label{eq:MSLL}
  M = 10\log\left(\frac{\max P_t(\theta_d, \phi_d)_{\theta_d, \phi_d \notin \psi}}{\max P_t(\theta_d, \phi_d)}\right),
\end{equation}
\noindent where $\psi$ represents the angular region of the mainlobe. 

\par Both the directivity $D$ and the maximum sidelobe level $M$ are determined by radiated power. The radiated power $P_t(\theta_d, \phi_d)$ is derived from the array factor $AF_t(\theta_d, \phi_d)$ at timeslot $t$, and is given by
\begin{equation}
  \label{eq:P_t}
  P_t(\theta_d, \phi_d) = |AF_t(\theta_d, \phi_d)|^2,
\end{equation}

\noindent where $AF_t(\theta_d, \phi_d)$ reflects the beamforming capabilities of the UAV-based VAA.

\subsection{Wind Field Model}
\label{subsec:wind_field_model}

\par The wind field model characterizes the impact of various wind conditions on the positional stability of UAVs during operations. The model considers three major types of wind field disturbances: constant, shear, and turbulent. Each wind type induces positional deviations in UAVs, described mathematically as follows:

\par In a constant wind field, the wind velocity $\mathbf{V_c}$ is uniform and can be expressed as follows:
\begin{equation}
  \mathbf{V_c} = V_c \cdot 
  \begin{pmatrix}
  \cos(\theta_c) \\ 
  \sin(\theta_c)
  \end{pmatrix},
\end{equation}

\noindent where $V_c$ is the wind speed magnitude and $\theta_c$ is the wind direction angle. The positional deviation of UAVs over timeslot $t$ due to constant wind is given by
\begin{equation}
  \label{eq:ConstantWindModel}
  \Delta \boldsymbol{\chi}_c(t) = \mathbf{V_c} \cdot t.
\end{equation}

\par Wind shear describes variations in wind speed with height. For a linear model, the wind speed $V_w(z)$ at height $z_w$ is given by
\begin{equation}
  V_w(z_w) = V_0 + k_w \cdot z_w,
\end{equation}

\noindent where $V_0$ is the wind speed at ground level and $k_w$ is the wind speed gradient. The positional deviation due to wind shear over timeslot $t$ is given by
\begin{equation}
  \label{eq:WindShearModel}
  \Delta \boldsymbol{\chi}_w(t) = \int_0^t (V_0 + k_w \cdot z_w) \, dt.
\end{equation}

\par Finally, atmospheric turbulence introduces random wind speed fluctuations, modeled by the Dryden turbulence model. The wind speed $V_{tur}(z_{tur})$ at height $z_{tur}$ is expressed as follows:
\begin{equation}
  V_{tur}(z_{tur}) = \frac{V_{\text{ref}}}{\ln(z_{tur}/z_0)}, \quad z_{tur} \geq z_0,
\end{equation}

\noindent where $V_{\text{ref}}$ is the wind speed at a reference height, and $z_0$ is the surface roughness length. Random fluctuations in wind speed at timeslot $t$ are given by
\begin{equation}
  V_{turb}(t) = \overline{V} + \sigma_V \cdot \sin(\omega_{turb} t + \phi_{turb}),
\end{equation}

\noindent where $\overline{V}$ is the average wind speed, $\sigma_V$ is the turbulence intensity, $\omega_{turb}$ is the turbulence frequency, and $\phi_{turb}$ is the random phase. The positional deviation due to turbulent wind is given by
\begin{equation}
  \label{eq:TurbulentWindModel}
  \Delta \boldsymbol{\chi}_{turb}(t) = \int_0^t V_{turb}(t) \, dt.
\end{equation}

\par In summary, the overall positional deviation of a UAV at timeslot $t$, combining all wind effects, is expressed as follows:
\begin{equation}
  \label{eq:WindFieldDisplacement}
  \Delta \boldsymbol{P}_{i,t} = 
  \begin{cases}
    \Delta \boldsymbol{\chi}_c(t), & \text{Constant Wind}, \\
    \Delta \boldsymbol{\chi}_w(t), & \text{Wind Shear}, \\
    \Delta \boldsymbol{\chi}_{turb}(t), & \text{Turbulent Wind}.
  \end{cases}
\end{equation}

\par This wind field model provides a detailed mathematical framework for evaluating the positional deviations of UAVs under different wind conditions. Based on this model, we can design robust UAV deployment strategies to maintain communication link stability.

\subsection{Problem Formulation}
\label{sec:problem_formulation_and_analysis}

\par In this work, we seek to address the challenges of UAV communication systems operating under positional disturbances caused by environmental factors such as wind. These disturbances introduce time-varying changes in system parameters, thereby significantly impacting communication performance. Specifically, the proposed system pursues two primary objectives, which are maximizing the directivity $D$ of the UAV-based VAA and minimizing the maximum sidelobe level $M$ to suppress interference and improve communication security. In the case of the positions of the UAVs being blown off course by the wind, we need to control the excitation current weights of the UAVs, thereby ensuring reliable and secure UAV communications in dynamic environments.

\par Accordingly, the decision variables within the considered system can be mathematically represented as a matrix $\boldsymbol{\Omega}$. Specifically, the matrix $\boldsymbol{\Omega} = \left\{ \omega_{i,t}^{\mathcal{U}} | i \in \mathcal{U}, t \in \mathcal{T} \right\}$ contains continuous variables representing the excitation current weights of the UAVs. Following this, the optimization problem can be formulated as follows:

\begin{subequations}
  \label{eq:formulation}
  \begin{align}
    {\underset{\boldsymbol{X} = \{\boldsymbol{\Omega}\} }{\text{max}}} \quad  & F=f\{D, M\},\\
    \text{s.t.} \quad \quad
    &\tau \in [\tau_{\text{min}}, \tau_{\text{max}}], \label{eq:const1}\\
    & t \in \mathcal{T}, \label{eq:const1}\\
    & 0 \leqslant \omega_{i,t}^{\mathcal{U}} \leqslant  1, \forall i \in \mathcal{U}, \forall t \in \mathcal{T}\label{eq:const2},\\
    & (x_{i,t}^{\mathcal{U}}, y_{i,t}^{\mathcal{U}}, z_{i,t}^{\mathcal{U}}) \in \mathbb{R}^{3}, \forall i \in \mathcal{U}, \label{eq:const3},\\
    & z_{i,t}^{\mathcal{U}} \geq z_{\text{min}}, \quad z_{i,t}^{\mathcal{U}} \leq z_{\text{max}}, \forall i \in \mathcal{U}, \label{eq:const4}
  \end{align}
\end{subequations}

\noindent where $\tau_{\text{min}}$ and $\tau_{\text{max}}$ represent the allowable time window for adjustment, and $\omega_{i,t}^{\mathcal{U}}$ denotes the excitation current weight for the $i$-th UAV, which ranges from 0 (UAV antenna is off) to 1 (UAV transmitting at full power). The physical position constraint is expressed in terms of 3D coordinates $(x_{i,t}^{\mathcal{U}}, y_{i,t}^{\mathcal{U}}, z_{n,t}^{\mathcal{U}})$, with bounds on the UAV altitude, which ensure that each UAV remains within a specified operational range $[z_{\text{min}}, z_{\text{max}}]$. 

\begin{table}[tb]
\centering
\caption{Summary of Key Notations}
\label{tab:notations}
\begin{tabularx}{\linewidth}{p{1.8cm}X}
\toprule
\textbf{Symbol} & \textbf{Description} \\ 
\midrule
$A_t$ & Advantage estimate in PPO-LA \\
$AF_t(\theta_d,\phi_d)$ & Array factor at timeslot $t$ \\
$\beta_1, \beta_2$ & Adam momentum parameters \\
$C_t$ & LSTM cell state \\
$c$ & Speed of light \\
$D$ & Directivity of VAA \\
$d_{min}$ & Minimum collision distance \\
$\Delta P_{i,t}$ & Total positional deviation of UAV $i$ at $t$ \\
$\Delta t$ & Time step size \\
$\Delta\chi_c(t)$ & Positional deviation from constant wind \\
$\Delta\chi_w(t)$ & Positional deviation from wind shear \\
$\Delta\chi_{turb}(t)$ & Positional deviation from turbulence \\
$\epsilon$ & PPO clip range hyperparameter \\
$f_c$ & Carrier frequency \\
$f_t, i_t, o_t$ & LSTM forget/input/output gates \\
$\gamma$ & Discount factor (implied in rewards) \\
$\Gamma$ & Ground BS \\
$H_{min}, H_{max}$ & Altitude range \\
$h_t$ & LSTM hidden state \\
$j$ & Imaginary unit $\sqrt{-1}$ \\
$k$ & Wavenumber $2\pi/\lambda$ \\
$k_w$ & Wind speed gradient \\
$L^{CLIP}$ & Clipped policy objective \\
$L^{VF}$ & Value function loss \\
$\lambda$ & Signal wavelength \\
$M$ & Maximum sidelobe level \\
$\hat{m}_t, \hat{v}_t$ & Bias-corrected moment estimates \\
$N$ & Number of UAVs in the swarm \\
$N_{pop}$ & Population size in PPO-LA \\
$\eta_t$ & Adaptive learning rate at $t$ \\
$\theta$ & Policy network parameters \\
$\theta_c$ & Wind direction angle \\
$\theta_d$ & Elevation angle $\in [0,\pi]$ \\
$\Omega$ & Excitation weight matrix $\{\omega_{i,t}\}$ \\
$\phi$ & Value network parameters \\
$\phi_d$ & Azimuth angle $\in [-\pi,\pi]$ \\
$\phi_{turb}$ & Random phase angle \\
$P_t(\theta_d,\phi_d)$ & Radiated power in direction $(\theta_d,\phi_d)$ \\
$\psi$ & Angular region of mainlobe \\
$\sigma_V$ & Turbulence intensity \\
$T$ & Total number of timeslots \\
$T_{max}$ & Maximum training iterations \\
$\mathcal{T}$ & Timeslot index set $\{1,2,\ldots,T\}$ \\
$\tau$ & Time window $[\tau_{min}, \tau_{max}]$ \\
$\mathcal{U}$ & Set of UAVs $\{1,2,\ldots,N\}$ \\
$V_0$ & Ground-level wind speed \\
$V_c$ & Constant wind velocity vector \\
$V_{ref}$ & Reference wind speed \\
$V_w(z_w)$ & Wind speed at height $z_w$ (shear model) \\
$V_{tur}(z_{tur})$ & Turbulent wind speed at height $z_{tur}$ \\
$V_{turb}(t)$ & Turbulent wind speed fluctuation \\
$\omega^U_{i,t}$ & Excitation current weight of UAV $i$ at $t$ \\
$\omega_{turb}$ & Turbulence frequency \\
$W_*, b_*$ & LSTM weight matrices and biases \\
$(x^U_{i,t}, y^U_{i,t}, z^U_{i,t})$ & 3D position of UAV $i$ at timeslot $t$ \\
$(x^\Gamma, y^\Gamma, z^\Gamma)$ & Fixed BS coordinates \\
$z_0$ & Surface roughness length \\
$z_{min}, z_{max}$ & UAV altitude bounds \\
\bottomrule
\end{tabularx}
\end{table}

\par The optimization problem focuses on maximizing communication quality by adjusting the excitation current weights $\omega_{i,t}^{\mathcal{U}}$ while adhering to operational constraints. Notably, the problem is non-convex due to the nonlinear interactions between current weight adjustments and CB performance. These nonlinearities are further exacerbated by the influence of wind-induced positional deviations on the UAVs.

%
\section{Algorithm Design}
\label{sec:algorithm}

\par In this section, we provide a comprehensive overview of the proposed algorithm, including the motivations for using DRL, Markov decision process (MDP) construction, the standard PPO algorithm, and the proposed PPO-LA.

\subsection{Motivations for Using DRL}

\par In this work, the optimization problem exhibits three critical characteristics that pose significant challenges for traditional approaches. First, the problem is inherently continuous, high-dimensional, and non-convex, as it involves real-time UAV positioning affected by wind field perturbations in high-altitude airspace. Second, the problem is stochastic and non-stationary, where the system states evolve unpredictably due to the dynamic nature of wind fields. Third, the problem requires sequential decision-making with delayed rewards, as the impact of positioning decisions on communication performance can only be evaluated over time. Traditional methods, such as convex optimization and evolutionary algorithms, are inherently limited in addressing these characteristics due to their discrete state-action space assumptions and inability to handle delayed feedback effectively.

\par Therefore, DRL emerges as a suitable solution precisely because its framework is designed to address these problem-specific challenges \cite{Sun2025Online, Liu2025LAMeTA}. The incorporation of deep neural networks enables the effective handling of continuous, high-dimensional state-action spaces through function approximation. Moreover, the online learning paradigm of DRL allows for adaptive policy optimization in stochastic and non-stationary environments, where the agent continuously refines its strategy based on real-time feedback. In addition, the capability of DRL to learn from experience through temporal difference learning makes it particularly suitable for problems with delayed rewards, thereby enabling the UAV system to optimize long-term communication performance rather than merely immediate gains. These inherent advantages of DRL directly correspond to and address the fundamental characteristics of the optimization problem.

\subsection{Construction of the MDP}

\par To apply DRL in the considered CB-based UAV communication system operating within a physical wind field, we need to transform our optimization problem as an MDP. In this case, we design the states, actions, and rewards of the MDP as follows:

\begin{itemize}
    \item \textbf{States ($s_t$):} The state space is designed to comprehensively capture both the system configuration and environmental dynamics that influence the performance of the considered CB-based UAV communication system operating within a physical wind field. Specifically, the state at time slot $t$ encompasses the three-dimensional spatial coordinates and excitation current weights of each UAV, alongside the wind field parameters characterized by velocity ($V_t$) and direction ($\theta_t$). This state representation ensures complete observability of both controllable system parameters and environmental disturbances. Mathematically, the state $s_t$ is designed as follows:
    \begin{equation}
        s_t = \{(x_{i,t}^{\mathcal{U}}, y_{i,t}^{\mathcal{U}}, z_{i,t}^{\mathcal{U}}), \omega_{i,t}^{\mathcal{U}}, V_t, \theta_t\}.
    \end{equation}

    \item \textbf{Actions ($a_t$):} The action space is constructed to facilitate direct control over the CB capabilities. At each time slot $t$, the action vector comprises the excitation current weights for all UAVs, \textit{i.e.},
    \begin{equation}
        a_t = \{\omega_{i,t}\}
    \end{equation}
    \noindent This action enables fine-grained control over the radiation pattern through weight adjustment, which allows the system to compensate for position perturbations induced by wind field disturbances while optimizing beamforming performance.

    \item \textbf{Rewards ($r_t$):} The reward function is carefully engineered to align the learning objective with the system performance requirements while incorporating necessary operational constraints. The designed reward function synthesizes multiple performance metrics into a scalar signal that guides the learning process toward optimal policies. The reward at timeslot $t$ is defined as follows:
    \begin{equation}
    \label{reward}
        r_t = D/M - p_0
    \end{equation}
    \noindent where $D$ represents the directivity, $M$ denotes the maximum sidelobe level, and $p_0$ is a penalty term for violating operational bounds. This reward function creates a balance between maximizing directivity and controlling sidelobe levels, while the penalty term ensures adherence to system constraints. The ratio $D/M$ is specifically chosen to normalize the competing objectives and provide a dimensionless reward signal that facilitates stable learning.
\end{itemize}

\par This MDP has a large state space and action space, which are difficult to handle. Therefore, we aim to propose an enhanced PPO algorithm to solve it. In the following, we first introduce the standard PPO algorithm.

\subsection{Standard PPO Algorithm}

\par PPO is widely regarded as one of the most effective DRL algorithms for continuous control tasks~\cite{Zhang2022}. The key advantage of PPO lies in the balance achieved between exploration and exploitation, which ensures stable policy updates while avoiding the instability often encountered in traditional policy gradient methods.

\par The core of PPO is the optimization of the policy through an objective function that strikes a balance between maximizing expected rewards and ensuring that updates to the policy do not deviate too significantly from the previous policy. The objective function for PPO is given by~\cite{Schulman2017}

\begin{align}
    L^{CLIP}(\theta) = \mathbb{E}_t \Big[ & \min \Big( \frac{\pi_\theta(a_t|s_t)}{\pi_{\theta_{\text{old}}}(a_t|s_t)} A_t, \nonumber \\
    & \text{clip}\left( \frac{\pi_\theta(a_t|s_t)}{\pi_{\theta_{\text{old}}}(a_t|s_t)}, 1 - \epsilon, 1 + \epsilon \right) A_t \Big) \Big],
    \end{align}

\noindent where $\pi_{\theta}(a_t|s_t)$ is the current policy, $\pi_{\theta_{\text{old}}}(a_t|s_t)$ is the old policy, $A_t$ is the advantage function, and $\epsilon$ is a small hyperparameter that controls the magnitude of policy updates. This objective ensures that large policy updates are penalized, which maintains stable learning dynamics while allowing for significant performance improvements.

\par However, the standard PPO algorithm still exhibits some limitations when applied to the considered system. First, the high dimensionality of both state and action spaces in our MDP poses considerable challenges for policy optimization, potentially leading to slow convergence and inefficient exploration. Second, the inherent dynamics of the wind field introduce substantial environmental uncertainties and system noise, which can destabilize the training process of standard PPO and compromise its robustness. These challenges are particularly critical in real-time UAV communication systems, where both convergence efficiency and policy stability directly impact system performance. The susceptibility of standard PPO to these issues may result in suboptimal policies that fail to maintain reliable communication performance under dynamic conditions. Therefore, we are motivated to enhance PPO to address these specific challenges.

\subsection{PPO-LA Algorithm}
\label{subsec:PPO-LA}

\par In this subsection, we propose a PPO-LA algorithm, which extends PPO by incorporating the LSTM structure and an adaptive learning rate method.

\subsubsection{Incorporation of LSTM Layers} 

\par The dynamic nature of wind field disturbances in the considered system poses significant challenges to the standard PPO algorithm. Specifically, the Markov assumption in standard PPO, which considers each decision step independently based solely on the current state, fails to capture the temporal correlations inherent in the system dynamics. This limitation leads to suboptimal policy learning and potential instability in the presence of time-varying disturbances.

\par To address these limitations, we propose to enhance the standard PPO framework by incorporating LSTM networks, which are specifically designed to capture temporal dependencies in sequential data. The LSTM-augmented policy network enables the consideration of both historical context and current observations in the decision-making process, which facilitates more robust and adaptive control strategies.

\par Specifically, the LSTM mechanism consists of a sophisticated gating architecture that regulates information flow through time. The core components include: 
\begin{align}
f_t &= \sigma(W_f \cdot [h_{t-1}, x_t] + b_f), \\
i_t &= \sigma(W_i \cdot [h_{t-1}, x_t] + b_i), \\
\tilde{C}_t &= \tanh(W_C \cdot [h_{t-1}, x_t] + b_C), \\
C_t &= f_t \times C_{t-1} + i_t \times \tilde{C}_t, \\
o_t &= \sigma(W_o \cdot [h_{t-1}, x_t] + b_o), \\
h_t &= o_t \times \tanh(C_t),
\end{align}
where $f_t$, $i_t$, and $o_t$ represent the forget, input, and output gates respectively, controlling the flow of information through the memory cell $C_t$. The hidden state $h_t$ serves as the final output, incorporating both current inputs and historical information through the carefully regulated memory mechanism.

\par Therefore, the integration of LSTM layers enables the capture of long-term dependencies in the system dynamics, which allows the policy to adapt to persistent patterns in wind field disturbances. 

\subsubsection{Adaptive Learning Rate Method}

\par The highly stochastic nature of the considered system introduces significant variations in gradient magnitudes during the training process. Standard PPO, which typically employs a fixed learning rate, struggles to efficiently navigate this varying optimization landscape. Specifically, a constant learning rate fails to adapt to the changing curvature of the loss surface, resulting in either slow convergence in shallow regions or instability in steep regions.

\par To address this limitation, we incorporate an adaptive learning rate mechanism based on the Adam optimizer into the PPO framework. This approach dynamically adjusts the effective learning rate by maintaining exponential moving averages of both the gradients and their second moments. The update rules are given by
\begin{align}
\hat{m}_t &= \frac{m_t}{1 - \beta_1^t}, \\
\hat{v}_t &= \frac{v_t}{1 - \beta_2^t}, \\
\theta_{t+1} &= \theta_t - \frac{\eta_t \cdot \hat{m}_t}{\sqrt{\hat{v}_t} + \epsilon},
\end{align}

\noindent where $\hat{m}_t$ and $\hat{v}_t$ represent the bias-corrected first and second moment estimates respectively, $\beta_1$ and $\beta_2$ are decay rates, and $\epsilon$ ensures numerical stability. This method enables automatic adjustment of step sizes based on the local geometry of the loss surface.

\par Therefore, the integration of adaptive learning rates enables efficient navigation of varying gradient landscapes, thereby automatically scaling step sizes based on local curvature, and thus improves robustness to different scales of gradients across network parameters.

\subsubsection{Main Steps and Computational Complexity of PPO-LA} 

\par
\begin{figure*}[t]
  \centering
  \includegraphics[width=\textwidth]{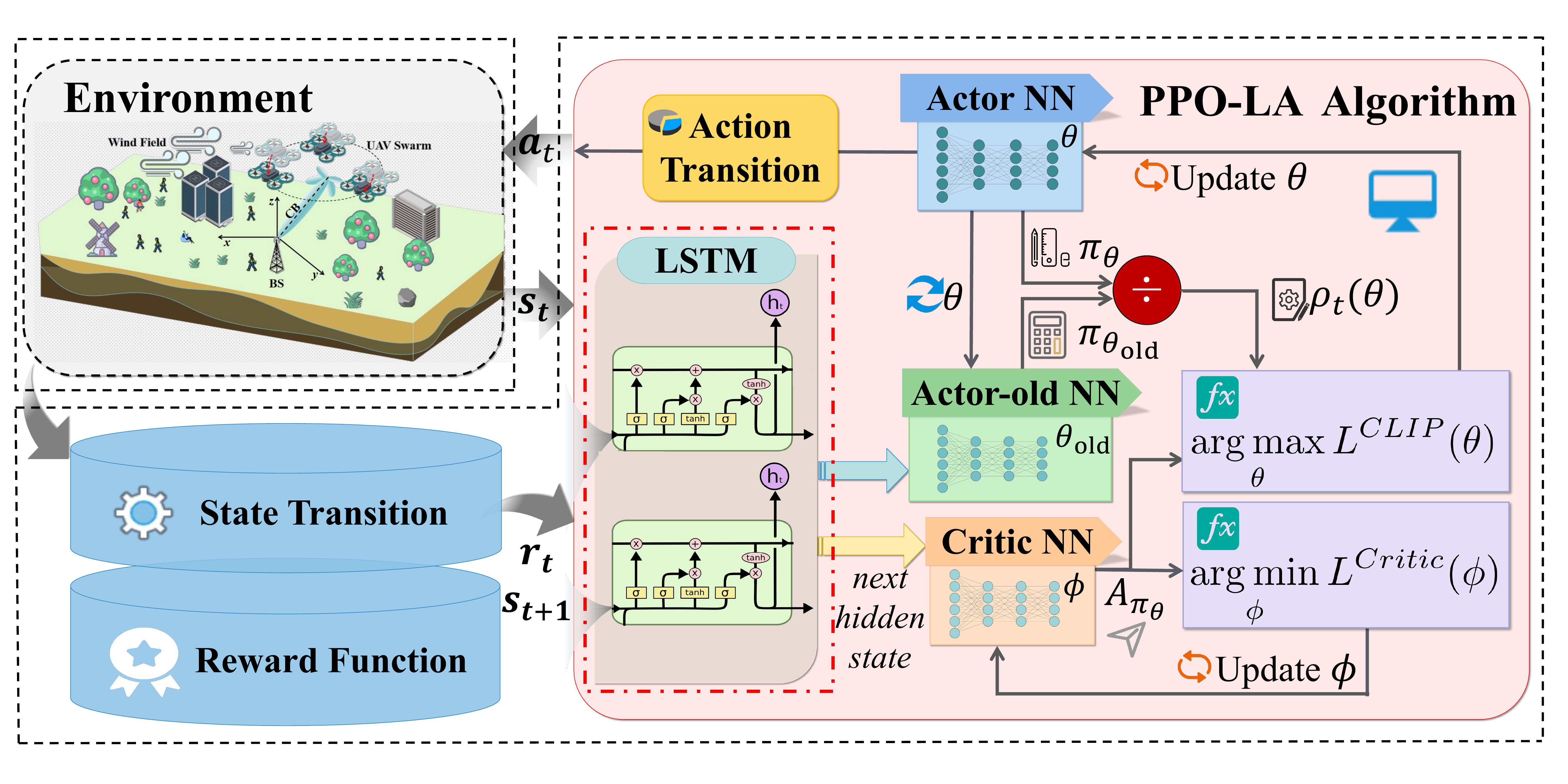}
  \caption{Framework of Recurrent PPO-LA.}
  \label{fig:algorithm-model}
\end{figure*}

 As illustrated in Fig.~\ref{fig:algorithm-model}, PPO-LA employs a three-network architecture: an actor network for policy learning with integrated LSTM layers to capture temporal dependencies, its corresponding actor-old network for computing probability ratios in the clipping mechanism, and a critic network for value function estimation. This architecture enables effective policy optimization while maintaining stability through the trust region constraint imposed by clipping.

\par The algorithm initializes the network parameters and episodic memory buffer for storing historical interactions. During each iteration, the agent interacts with the environment following the current policy $\pi_\theta$, where the actor network processes the current state $s_t$ through its LSTM layers to generate action $a_t$. The environment then transitions to state $s_{t+1}$ and yields reward $r_t$, which is designed to reflect the communication quality and system performance objectives.

\par Algorithm~\ref{alg:PPO-LA} outlines the main steps of PPO-LA. The training process consists of iterative episodes of environment interaction and policy optimization. Within each episode, the algorithm performs the following steps:

\begin{enumerate}
\item Collects trajectory data $\mathcal{D} = \{(s_t, a_t, r_t, s_{t+1})\}_{t=1}^T$ and computes advantage estimates
\[
A_t = \sum_{t'=t}^T \gamma^{t'-t} \Biggl( r_{t'} - V_\phi(s_{t'}) \Biggr) + \gamma^T V_\phi(s_T)
\]
by executing the current policy until the experience replay buffer reaches the specified batch size. 
\item Updates the policy network parameters $\theta$ by maximizing the clipped objective function $L^{CLIP}(\theta)$. 
\item Updates the value network parameters $\phi$ by minimizing the value function loss $L^{VF}(\phi) = \mathbb{E}_t \left[ \left( V_\phi(s_t) - R_t \right)^2 \right]$.
\item Updates the actor-old network parameters to match the current actor network. 
\end{enumerate}

\par This iterative process continues until convergence, which allows the policy to adaptively improve its performance in the dynamic UAV communication environment while maintaining training stability through the clipping mechanism and temporal context through the LSTM layers.

\subsection{Computational Complexity of PPO-LA} 

\par Subsequently, we analyze the computational complexity of the PPO-LA as follows:

\paragraph{Training Phase}
\par The computational complexity of PPO-LA during the training phase is 
$
O(2|\theta_\pi| + |\theta_v| + M_aN_aT|\theta_\pi| + M_aN_aV + M_aN_a(2|\theta_\pi| + |\theta_v|)),
$
\noindent where $M_a$ is the number of training episodes, $N_a$ is the number of steps per episode, $T$ is the number of timeslots required for action sampling, $V$ is the complexity of interacting with the environment, $|\theta_\pi|$ is the number of parameters in the actor network, and $|\theta_v|$ is the number of parameters in the critic network \cite{Zhang2025a}.

\par Specifically, this complexity can be broken down into the following parts:

\begin{itemize}
    \item \textbf{Network Initialize:} This phase involves the initialization of the policy and value network parameters. The computational complexity for this phase is expressed as $O(2|\theta_\pi| + |\theta_v|)$.
    
    \item \textbf{Action Sampling:} This phase entails generating actions according to the current policy $\pi_\theta$. The complexity of generating actions for each timeslot is $O(|\theta_\pi|)$. Since this is done for $T$ timeslots per episode and $M$ episodes, the total complexity for action sampling is $O(M_aN_aT|\theta_\pi|)$.
    
    \item \textbf{Replay Buffer Collection:} The complexity of collecting state transitions in the replay buffer is $O(M_aN_aV)$. The environment interaction happens at each step for each episode, and $V$ typically accounts for any additional computations needed during interaction, such as state and reward calculations.
    
    \item \textbf{Network Update:} The update phase consists of frequent updates of the policy network and less frequent updates of the value network. The complexity of updating both networks is $O(M_aN_a(2|\theta_\pi|) + M_aN_a|\theta_v|)$, where the factor $2|\theta_\pi|$ accounts for the need to compute both the policy loss and the value loss for each timeslot, and $|\theta_v|$ represents the complexity of updating the value network.
\end{itemize}

\par Subsequently, the space complexity during the training phase is
$
O(|\theta_\pi| + |\theta_v| + D_a(2|s| + |a| + 1)), 
$
where $D_a$ is the size of the replay buffer, $|s|$ is the dimensionality of the state space, $|a|$ is the dimensionality of the action space.

This space complexity accounts for the storage of network parameters, as well as the data structures required to maintain the replay buffer, which stores tuples of states, actions, rewards, and next states.

\paragraph{Execution Phase}
\par During the execution phase, the computational complexity of PPO-LA is $O(M_aN_aT|\theta_\pi|)$. This complexity is primarily due to action selection according to the current policy.

\par The space complexity during the execution phase is $O(|\theta_\pi|)$, since the policy network parameters need to be stored in memory for action selection.

\begin{algorithm}[t]
\caption{PPO-LA Algorithm}
\label{alg:PPO-LA}
\KwIn{Population size $N_{\mathrm{pop}}$, number of maximum iterations $T_{\mathrm{max}}$, objective function $f$.} 
\tcc{\textbf{Initialization stage}}
\textbf{Initialize} policy network parameters $\theta$, actor old network parameters $\theta_{\text{old}}$, value network parameters $\phi$\ and episodic memory $\mathcal{B}$\;
\tcc{\textbf{Data Collection Stage}}
\For{each Episode}{
    Reset the environment and initialize state $s_t$\;

    \tcc{\textbf{Trajectory Generation}}
    \For{each Time slot}{
        Collect trajectory data $\mathcal{D}$;
        
        Compute advantage estimates $A_t$;

        Calculated $r_t$ according to Eq. (\ref{reward});
        
        Store transition $(s_t, a_t, r_t, s_{t+1})$ in $\mathcal{B}$\;

    \tcc{\textbf{Policy Optimization Stage}}
    \If{size of buffer $\mathcal{B}$ == batch size}{
        \For{each update step}{
            Sample a batch of transitions from $\mathcal{B}$\;
            
            \tcp{Update policy with clipping mechanism}
            Update policy network parameters $\theta$ by maximizing $L^{CLIP}(\theta)$\;
            
            \tcp{Update value function}
            Update value network parameters $\phi$ by minimizing $L^{VF}(\phi)$;
            
        }
    }
    \tcp{Network synchronization}
    $\theta_{\text{old}} \gets \theta$\;
    Clear buffer $\mathcal{B} \gets \emptyset$\;
    }
    \textbf{Check termination condition}\;
}
\KwOut{Optimized policy parameters $\theta^*$ for deployment.}
\end{algorithm}

%
%
\section{Simulation Results and Analysis} %
\label{sec:simulation_results_and_analysis}

\par In this section, we present comprehensive simulation results to evaluate the effectiveness of the proposed approach.

%
\subsection{Simulation Setups}
\label{ssec:simulation_setups}


\par We investigate two distinct UAV swarm configurations to evaluate algorithmic scalability: (i) a small-scale swarm with $8$ UAVs, and (ii) a large-scale swarm with $16$ UAVs, operating within a $50\times50\times50~\mathrm{m}^3$ airspace. This dual-scenario approach validates the robustness of our method across varying environmental dimensions. Key physical parameters include: minimum collision distance $d_{\min} = 0.5~\mathrm{m}$, and altitude range $H_{\min} = 30~\mathrm{m}$ to $H_{\max} = 50~\mathrm{m}$. 

\par The communication system adopts the following specifications: carrier frequency $f_c = 3~\mathrm{GHz}$ with corresponding wavelength $\lambda = 0.1~\mathrm{m}$, where $c = 3\times10^8~\mathrm{m/s}$ is the speed of light.  The simulation runs for $T = 100~\mathrm{s}$ with discrete time steps $\Delta t = 0.5~\mathrm{s}$. The initial position and excitation current of UAV swarm are under optimal CB communication conditions. Other parameters related to communication follow reference \cite{Wang2023}.

\par For comparison, we utilize deep deterministic policy gradient (DDPG) \cite{lillicrap2015continuous}, deep Q-learning (DQN) \cite{mnih2015human}, soft actor-critic (SAC) \cite{haarnoja2018soft}, and standard PPO \cite{schulman2017proximal} as the benchmark algorithms.

\begin{figure*}
\centering
\subfloat[]{
\includegraphics[width=0.28\linewidth]{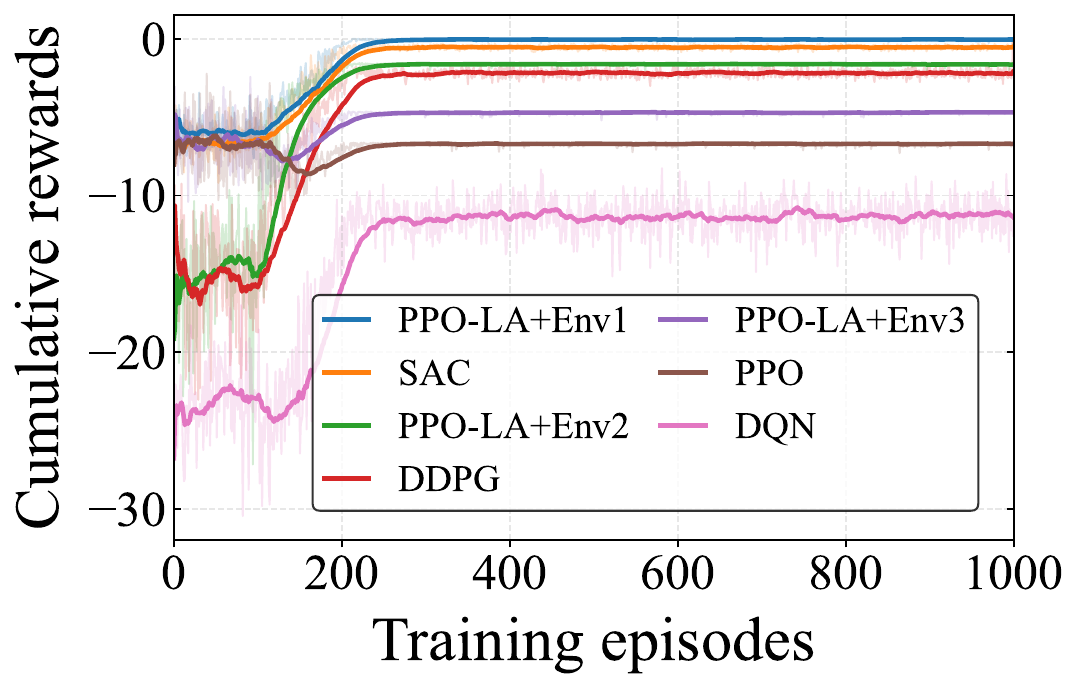}\label{subfig:1}}
\subfloat[]{
\includegraphics[width=0.44\linewidth]{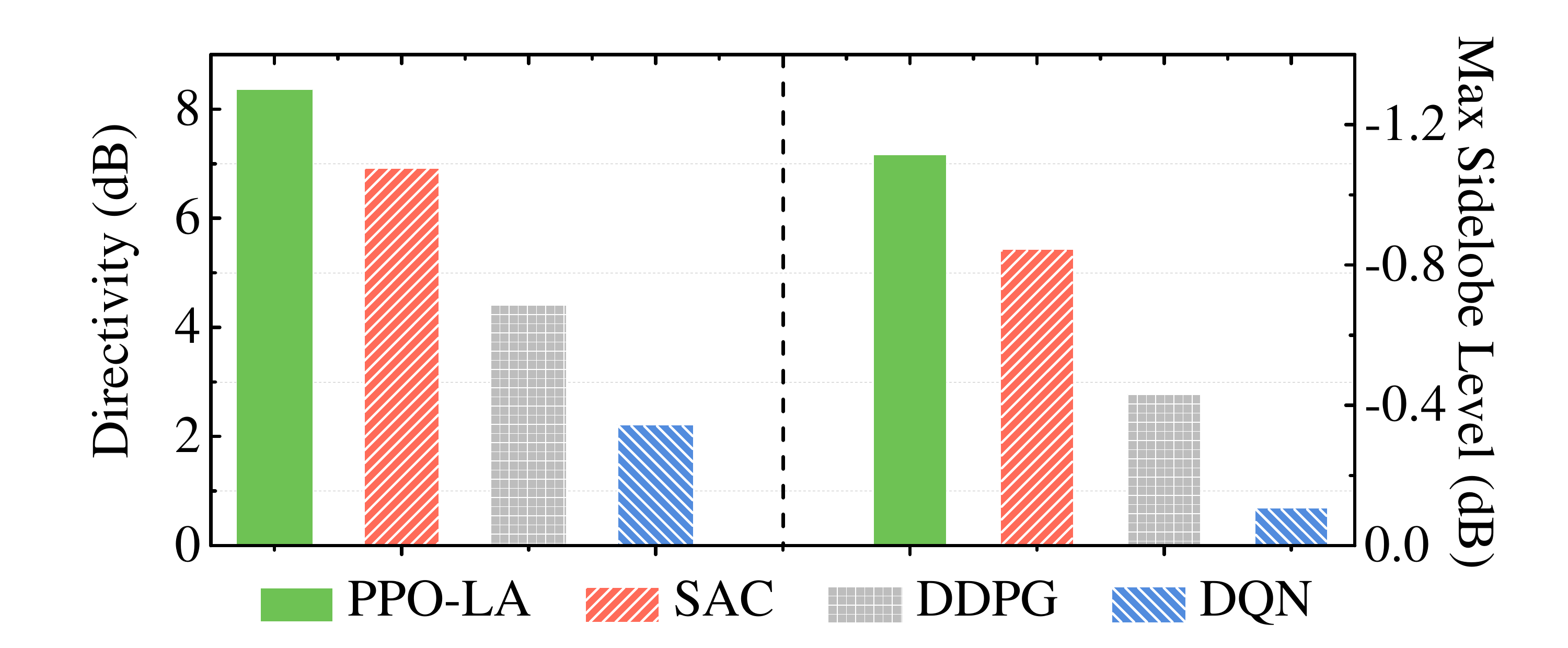}\label{subfig:2}}
\subfloat[]{
\includegraphics[width=0.28\linewidth]{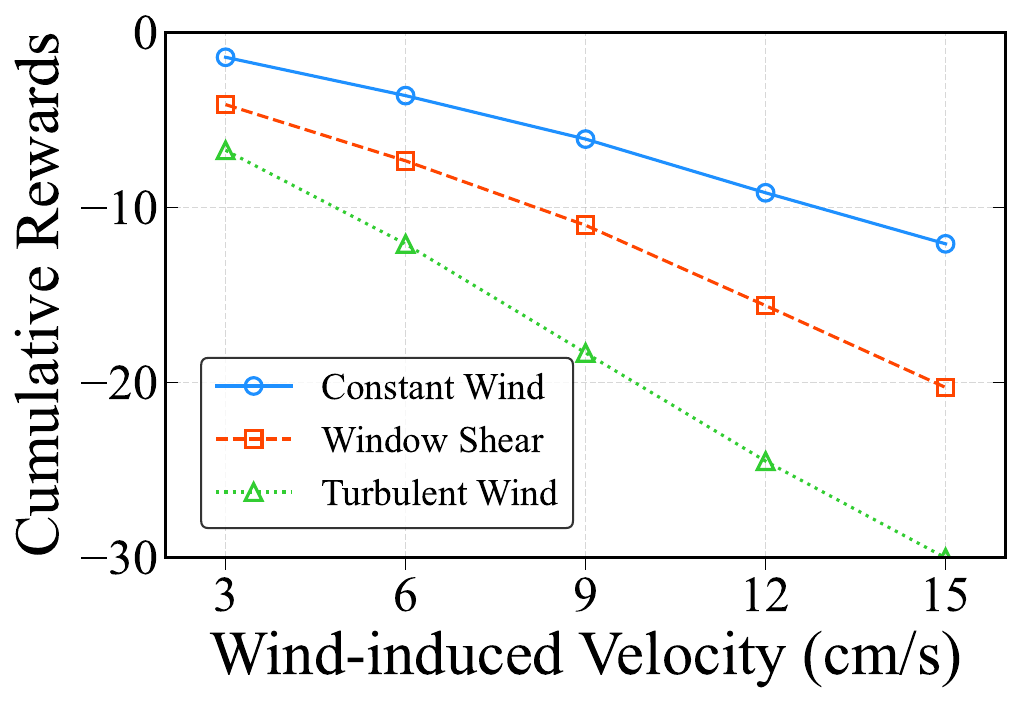}\label{subfig:3}}
\

\caption{Simulation results with 8 UAVs. (a) Cumulative rewards training curve. (b) Dual-objective CB optimization analysis. (c) Quantitative impact of wind speed on communication performance. }
\label{fig:Robustness-results}
\end{figure*}

%
\subsection{Results and Analyses}
\label{ssec:visualization_results}

\par Fig. \ref{fig:Robustness-results}\subref{subfig:1} illustrates the cumulative rewards across episodes for PPO-LA compared to benchmark and original algorithms under various wind conditions. It can be observed, while all algorithms initially demonstrate unstable rewards due to boundary violations and wind field disturbances, PPO-LA achieves significantly higher reward values and faster convergence once stabilized. This indicates that the proposed PPO-LA algorithm exhibits superior learning efficiency and robustness in optimizing the CB performance of UAV swarms under wind field disturbances. This superior performance can be attributed to the fact that the LSTM structure effectively captures temporal dependencies in wind field patterns, which enables anticipatory weight adjustments within acceptable latency constraints. Moreover, the adaptive learning rate mechanism facilitates a more efficient exploration-exploitation balance, particularly valuable when confronting highly dynamic turbulent wind conditions.

\par Fig. \ref{fig:Robustness-results}\subref{subfig:2} presents the comparative performance of different algorithms in optimizing directivity $D$ and maximum sidelobe level $M$ simultaneously. The results demonstrate, PPO-LA consistently achieves higher directivity while maintaining lower sidelobe levels compared to benchmark algorithms across different wind conditions. This indicates that the proposed algorithm effectively balances the multi-objective optimization problem, which maintains communication quality even under challenging environmental disturbances. This effectiveness stems from our dual-phase reward mechanism (Eq. (\ref{eq:formulation})) dynamically prioritizes between directional gain maximization and interference suppression based on current wind conditions, which enables real-time adaptation of excitation current weights to compensate for positional errors. 
\begin{figure}[htbp]
\centering
\subfloat[]{
\includegraphics[width=0.85\linewidth]{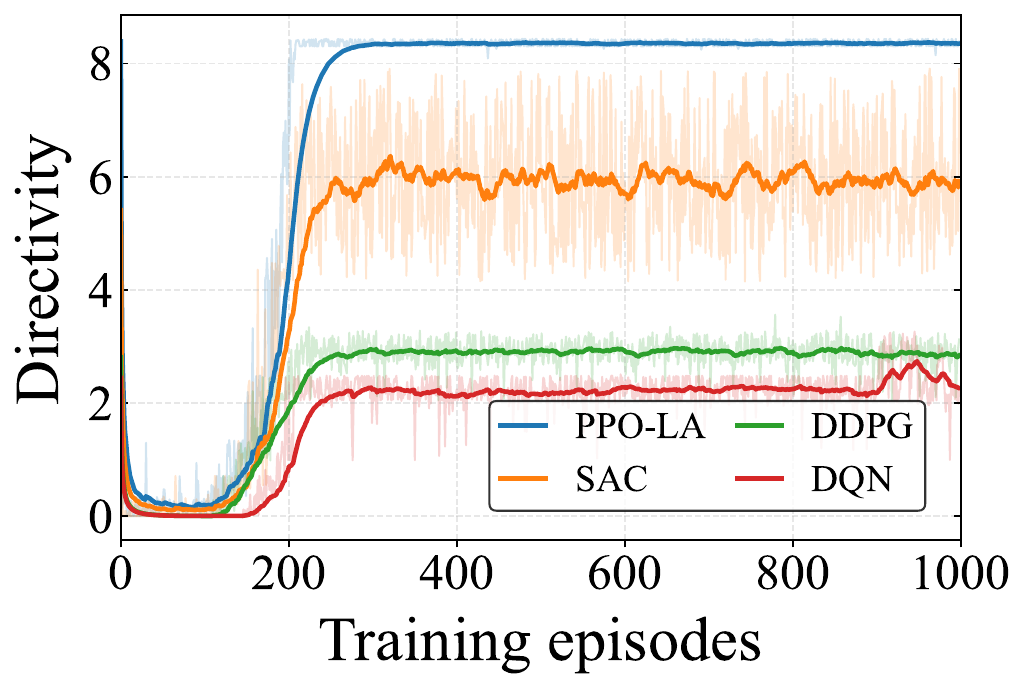}\label{subfig:11}}

\subfloat[]{
\includegraphics[width=0.85\linewidth]{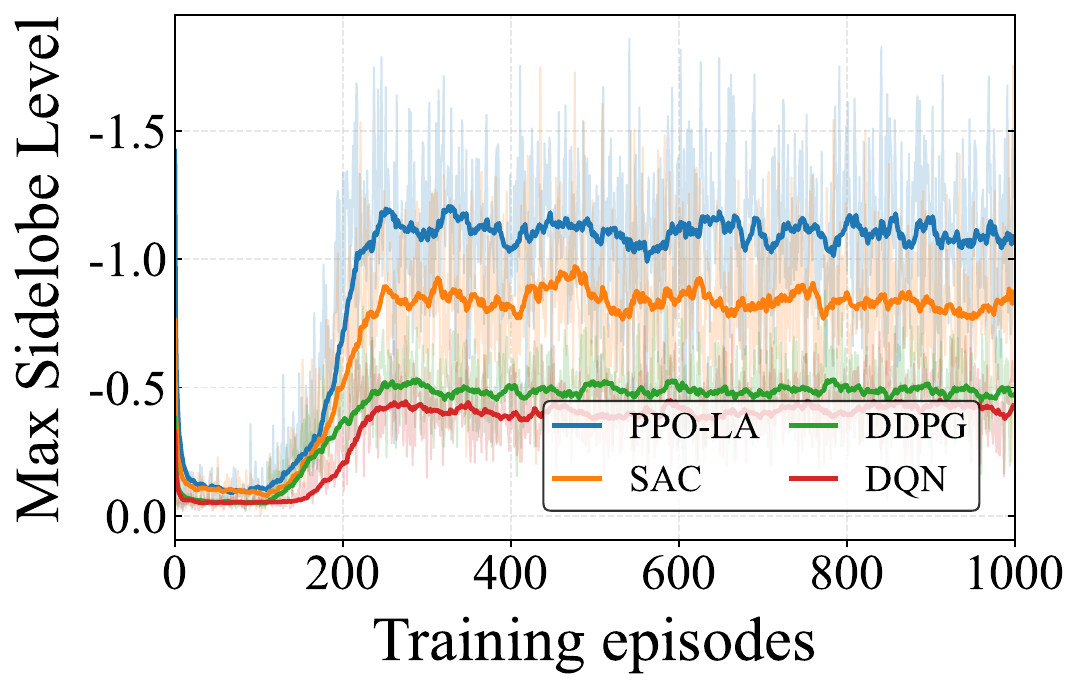}\label{subfig:22}}

\caption{Simulation results with 8 UAVs. (a) Directivity Training curve. (b) Max Sidelobe Level Training curve.}
\label{fig:44-results}
\end{figure}

\begin{figure}[htbp]
\centering
\subfloat[]{
\includegraphics[width=0.85\linewidth]{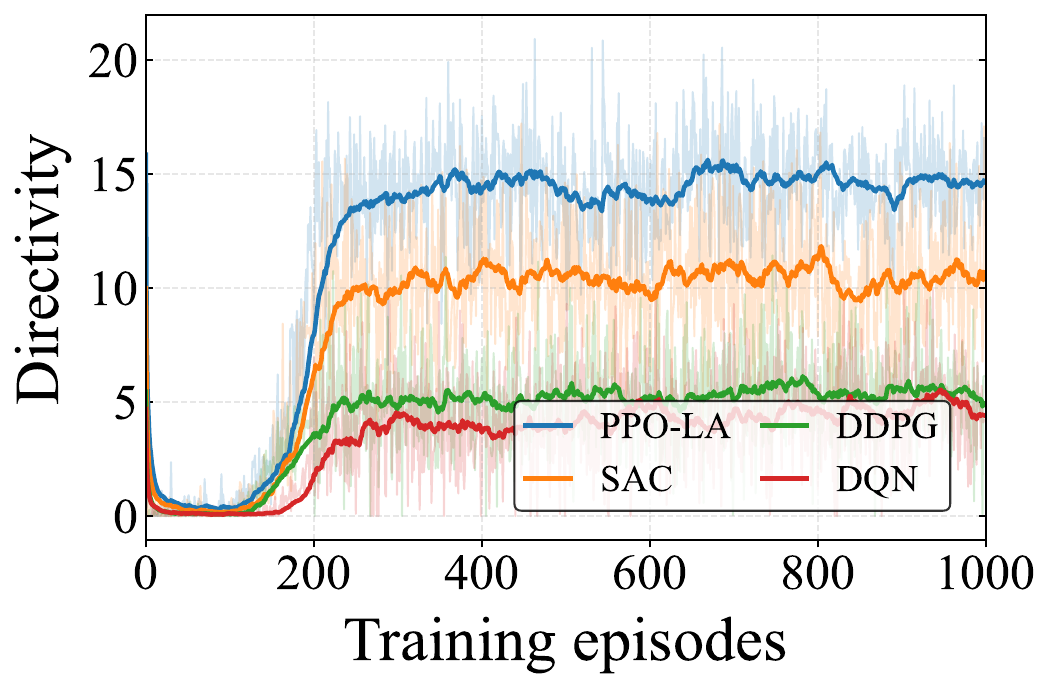}\label{subfig:33}}

\subfloat[]{
\includegraphics[width=0.85\linewidth]{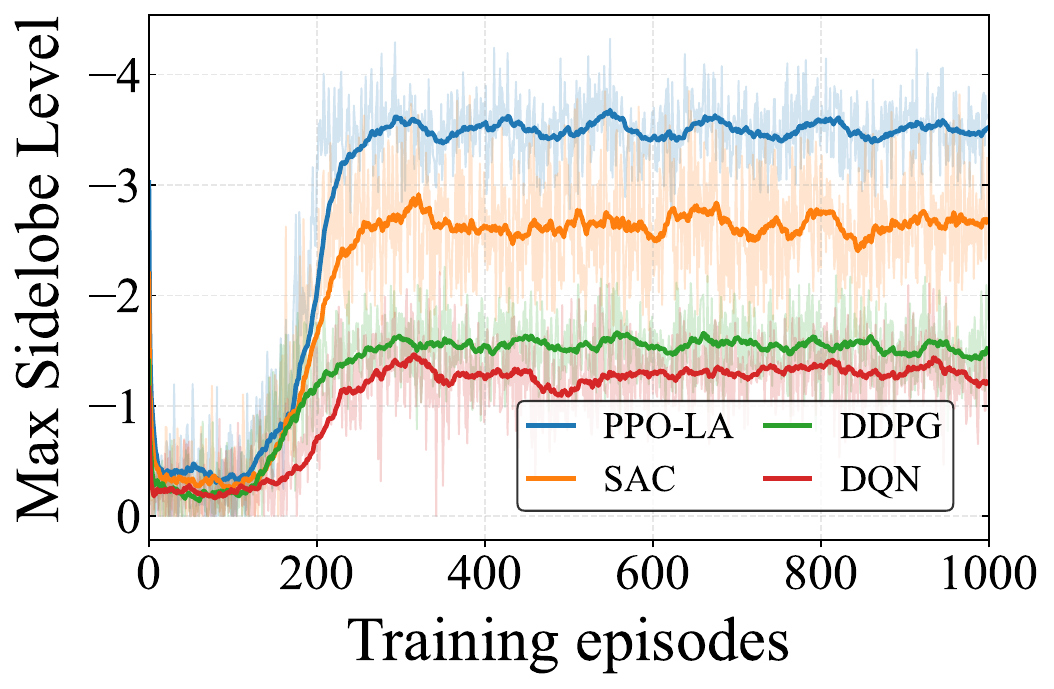}\label{subfig:44}}

\caption{Simulation results with 16 UAVs. (a) Directivity Training curve. (b) Max Sidelobe Level Training curve.}
\label{fig:55-results}
\end{figure}

\par Fig. \ref{fig:Robustness-results}\subref{subfig:3} quantifies the relationship between CB performance metrics and wind speed across different wind field types. The analysis reveals that performance degradation accelerates with increasing wind speed, with turbulent wind fields causing more severe deterioration than constant or shear wind conditions at equivalent speeds. This indicates that the proposed PPO-LA algorithm, while effectively mitigating performance loss across all conditions, demonstrates particular advantage in turbulent environments where conventional approaches fail dramatically. This superior performance in turbulent conditions can be attributed to the fact that turbulent wind introduces complex spatial-temporal patterns of UAV displacement, which the LSTM component of PPO-LA can model and predict, which allows preemptive compensatory adjustments to excitation current weights before significant performance degradation occurs.

\par Fig. \ref{fig:44-results} provides a detailed demonstration of the convergence processes of both directivity $D$ and maximum sidelobe level $M$ for PPO-LA and benchmark algorithms under wind field disturbances. This figure serves as an extended analysis of Fig. \ref{fig:Robustness-results}\subref{subfig:3}. It can be observed that during the initial phase (0-200 iterations), all algorithms exhibit significant performance fluctuations. Notably, during the first few iterations, both $D$ and $M$ maintain relatively high values, with the average initial $D$ reaching 8.38 dB and $M$ remaining around -1.07 dB. This phenomenon only occurs at the beginning of algorithm initialization and quickly stabilizes as the number of iterations increases. This situation primarily stems from two factors: (1) the cumulative effects of wind field disturbances have not yet fully manifested, and (2) the agents have not yet developed effective anti-disturbance strategies. 

\par As the training progresses, the algorithms begin to demonstrate distinct performance differentiation. PPO-LA exhibits remarkable advantages: first, its convergence speed is faster than benchmark algorithms, which benefits from the rapid capture of spatiotemporal wind field characteristics by the LSTM module; second, during the stable phase, PPO-LA maintains a $D$ value (8.25±0.15 dB) that is higher than the best benchmark algorithm, while also demonstrating superior performance in $M$. This simultaneous optimization validates the effectiveness of the dual-phase reward mechanism, which further demonstrates capability of PPO-LA to achieve higher directivity $D$  while maintaining lower maximum sidelobe level $M$.

\par Fig. \ref{fig:55-results} presents the performance comparison of directivity $D$ and maximum sidelobe level $M$ between PPO-LA and benchmark algorithms in 16-UAV formations, which validates the effectiveness of algorithm in larger-scale UAV swarms. The experimental results reveal that the stabilized D value increases to 15dB (versus 8.25dB in 8-UAV formations), while M value correspondingly changes to -3.5dB, which confirms the correlation between the number of UAVs and both directivity $D$ and maximum sidelobe level $M$. In particular, the fluctuation range of $D$ values expands significantly. This situation primarily stems from the error accumulation effect in large UAV arrays. Specifically, individual UAV positioning errors become magnified through array factor multiplication,thereby causing more pronounced communication quality fluctuations and illustrating the enhanced cooperative control challenges in scaled-up systems under wind disturbances.

%
%
\section{Conclusion} %
\label{sec:conclusion}

\par In this paper, we have investigated the impact of wind-induced position errors on CB performance in UAV swarm communications and proposed a novel DRL solution to address this critical challenge. First, we have proposed a comprehensive framework that models the performance degradation of VAA under three distinct wind field conditions (constant, shear, and turbulent), which quantifies the negative effects on directivity $D$ and maximum sidelobe level $M$. Subsequently, we have formulated a long-term real-time optimization problem and developed the PPO-LA algorithm to dynamically adjust the excitation current weights of UAVs without requiring extensive prior training for specific wind field conditions. The simulation results have demonstrated that the proposed PPO-LA algorithm significantly outperforms conventional methods in terms of convergence speed, stability, and overall communication performance under various scenarios of wind disturbances.


\bibliographystyle{IEEEtran}
\bibliography{myref}

\end{document}